\documentclass[article,aps,prl,twocolumn,superscriptaddress,english,longbibliography]{revtex4-2}

\usepackage{bm,amssymb,amsmath,dsfont,mathrsfs,amstext,latexsym,physics,relsize}
\usepackage{graphicx}

\usepackage[usenames,dvipsnames]{xcolor}
\usepackage[colorlinks=true,citecolor=blue,urlcolor=blue,linkcolor=blue]{hyperref}
\usepackage[all]{hypcap}

\newcommand{\figpanel}[2]{\hyperref[#1]{\ref{#1}#2}}

\begin{document}

\title{Quantum Quenches from the Critical Point: Theory and Experimental Validation in a Trapped-Ion Quantum Simulator}

\author{Chen-Xu Wang}
\affiliation{Laboratory of Quantum Information, University of Science and Technology of China,
Hefei 230026, China}
\affiliation{Anhui Prvince Key Laboratory of Quantum Network, University of Science and Technology of China, Hefei 230026, China}
\affiliation{CAS Center For Excellence in Quantum Information and Quantum Physics,
University of Science and Technology of China, Hefei 230026, China }
\email{wcx1221@mail.ustc.edu.cn}

\author{Andr\'as Grabarits\href{https://orcid.org/0000-0002-0633-7195}{\includegraphics[scale=0.05]{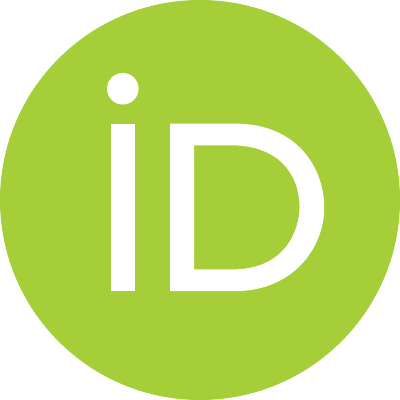}}}
\email{andras.grabarits@uni.lu}
\affiliation{Department  of  Physics  and  Materials  Science,  University  of  Luxembourg,  L-1511  Luxembourg, Luxembourg}

\author{Jin-Ming Cui}
\email{jmcui@ustc.edu.cn}
\affiliation{Laboratory of Quantum Information, University of Science and Technology of China, Hefei 230026, China}
\affiliation{Anhui Prvince Key Laboratory of Quantum Network, University of Science and Technology of China, Hefei 230026, China}
\affiliation{CAS Center For Excellence in Quantum Information and Quantum Physics,
University of Science and Technology of China, Hefei 230026, China }
\affiliation{Hefei National Laboratory, University of Science and Technology of
China, Hefei 230088, China }

\author{Hua-Bi Zeng}
\email{zenghuabi@hainanu.edu.cn}
\affiliation{Center for Theoretical Physics, Hainan University, Haikou 570228, China}

\author{Yun-Feng Huang}
\email{hyf@ustc.edu.cn}
\affiliation{Laboratory of Quantum Information, University of Science and Technology of China, Hefei 230026, China}
\affiliation{Anhui Prvince Key Laboratory of Quantum Network, University of Science and Technology of China, Hefei 230026, China}
\affiliation{CAS Center For Excellence in Quantum Information and Quantum Physics,
University of Science and Technology of China, Hefei 230026, China }
\affiliation{Hefei National Laboratory, University of Science and Technology of
China, Hefei 230088, China }

\author{Chuan-Feng Li}
\email{cfli@ustc.edu.cn}
\affiliation{Laboratory of Quantum Information, University of Science and Technology of China, Hefei 230026, China}
\affiliation{Anhui Prvince Key Laboratory of Quantum Network, University of Science and Technology of China, Hefei 230026, China}
\affiliation{CAS Center For Excellence in Quantum Information and Quantum Physics,
University of Science and Technology of China, Hefei 230026, China }
\affiliation{Hefei National Laboratory, University of Science and Technology of
China, Hefei 230088, China }

\author{Adolfo del Campo\href{https://orcid.org/0000-0003-2219-2851}{\includegraphics[scale=0.05]{orcidid.pdf}}}
\email{adolfo.delcampo@uni.lu}
\affiliation{Department  of  Physics  and  Materials  Science,  University  of  Luxembourg,  L-1511  Luxembourg, Luxembourg}
\affiliation{Donostia International Physics Center,  E-20018 San Sebasti\'an, Spain}

\date{today}

\begin{abstract}
We investigate quantum quenches starting from a critical point and experimentally probe the associated defect statistics using a trapped-ion quantum simulator of the transverse-field Ising model. The cumulants of the defect number distribution exhibit universal scaling with quench depth, featuring Gaussian behavior at leading order and systematic subleading corrections. Our results are in excellent agreement with both exact and approximate theoretical predictions, establishing quench-depth scaling as a powerful and precise experimental benchmark for nonequilibrium quantum critical dynamics.
\end{abstract}

\maketitle

A fundamental objective in condensed matter and statistical physics is to determine the conditions under which universal behavior emerges in quantum many-body systems. In equilibrium settings, such universality, particularly near continuous phase transitions, is captured by the Ginzburg–Landau–Wilson paradigm, which classifies systems according to symmetries and spatial dimensionality via critical exponents~\cite{Sachdev2011}.

Out of equilibrium, the emergence of universality is more subtle, but can still be captured by the Kibble–Zurek (KZ) mechanism~\cite{Kibble1976,Zurek1985,Zurek1996,DZ14}. This framework describes systems slowly driven across continuous phase transitions in a finite quench time $\tau_Q$. As the control parameter $g$ approaches its critical value $g_c$, the relaxation time diverges as $\tau \sim |\epsilon|^{-z\nu}$ as a function of the reduced parameter $\epsilon = (g - g_c)/g_c$, signaling a breakdown of the adiabatic dynamics near the critical point. For a linearized quench $\epsilon=t/\tau_Q$, the effective nonequilibrium relaxation time is given by the so-called freeze-out time $\hat{t}$, identified by the condition $\tau(\hat{t}) = \hat{t}$, resulting in $\hat{t} \sim (\tau_0 \tau_Q^{z\nu})^{1/(1+z\nu)}$. This in turn sets the nonequilibrium KZ correlation length scale $\hat{\xi} \sim |\epsilon(\hat{t})|^{-\nu}$, which determines the typical size of the domains created during the transition. As a result, the density of point-like topological defects in $d$ spatial dimensions scales universally as $n \sim \hat{\xi}^{-d} \sim \tau_Q^{-d\nu/(1 + z\nu)}$.

In the quantum regime, the KZ mechanism provides an effective framework for describing defect generation in slowly driven quantum phase transitions (QPTs) \cite{Damski05,Dziarmaga05,Zurek05,DZ14} and has become a widely used benchmark for quantum simulators and quantum computers \cite{Cui2016,Gardas2018,Weinberg20,Cui2020,Bando20,King22,Duan23,Ali2024,Miessen24,King25,Zhang2025}.  Although KZ scaling accurately predicts defect formation near the adiabatic limit, in which few excitations are produced \cite{Zurek05}, it fails to capture dynamics under moderate or fast quenches. Recent studies have revealed a universal departure from the KZ behavior in this regime, where the defect density becomes governed by the final quench depth $\epsilon_f$ rather than the quench rate~\cite{Huabi2023FastQuench,Xia23,TFIM_FastQuench_2023,xia_Disk2024,Grabarits_Arbitrary2025}. Specifically, below a characteristic quench timescale $\tau_Q \lesssim \epsilon_f^{-(z\nu + 1)}$, the defect density saturates and scales as $n \sim \epsilon_f^{d\nu}$, becoming independent of $\tau_Q$. 

This breakdown of KZ scaling for the average defect number has been experimentally observed across various platforms, including confined ion chains~\cite{EH13,Ulm13,Pyka13}, ion-trap quantum simulators~\cite{Rao25}, ultracold Bose and Fermi gases driven through superfluid transitions~\cite{Ko2019,Donadello2016,Goo2021,Goo2022}, and superconducting rings~\cite{delCampo2010}. However, such studies focus on the conventional KZ setting in which a system is driven from one high-symmetry phase to a broken-symmetry phase, without addressing the universality of moderate and fast quenches. 

A particularly relevant scenario arises when the quench protocol begins in the vicinity of the critical point, for example, in the frozen regime. This naturally occurs in experiments with limited tunability of control fields and restricted access to the range of system parameters.  A qualitative theoretical insight based on finite-size scaling has been presented for slow quenches~\cite{Chandran2012}, compatible with KZ. In contrast, the fast-quench regime remains uncharted.  In such a scenario, one can expect richer phenomenology and improved analytical tractability, enabling a deeper understanding of the underlying dynamics.

In this Letter, we report the experimental study of the full counting statistics of defects generated in fast quantum quenches across QPTs, using a trapped-ion quantum simulator~
that realizes the one-dimensional transverse-field quantum Ising model (TFQIM). Starting from the quantum critical point, we uncover new universal scaling relations that depend on the final quench depth.
We find that both the average defect pair number and its variance scale approximately linearly with the final quench depth, while the third cumulant exhibits subleading quadratic corrections.  Moreover, in the slow-driving regime, we demonstrate that all cumulants follow a Kibble–Zurek-like universal scaling even when the quench commences at criticality. Finally, we show that, in both cases, defect pairs preserve their sub-Poissonian behavior, and defect numbers exhibit super-Poissonian statistics. These experimental findings are supported by exact results in the fast-quench limit and by analytical approximations combined with numerical simulations in the slow-driving regime.

\emph{TFQIM and experimental set-up.} 
The TFQIM provides a convenient testbed of both the equilibrium properties and the non-equilibrium dynamics of QPTs \cite{Sachdev2011,Suzuki2012Quantum,Dziarmaga05,Zurek05}. Its Hamiltonian is given by
\begin{eqnarray}
    \mathcal{\hat H}(t)=-J\sum_{m=1}^N\,\left(\hat\sigma^z_{m}\hat\sigma^z_{m+1}+g\hat\sigma^x_m\right),
\end{eqnarray}
where $\hat\sigma^{x,z}_m$ are the Pauli operators on the $m$-th spin with periodic boundary conditions, 
and we choose even $N$. Here, $g$ controls the strength of the transverse field and $J\equiv1$ fixes the energy scale, favoring ferromagnetic order.  The second-order critical points $g_c=\pm1$ separate the ferromagnetic $(|g|<1)$ and paramagnetic phases $(|g|>1)$. 
The TFQIM Hamiltonian 
can be mapped to the direct sum of independent two-level systems (TLSs) 
$\mathcal{\hat H}(t)=\sum_{k>0}\hat\varphi^\dagger_k\left[(g-\cos k)\sigma^z+\sin k\sigma^x\right]\hat\varphi_k=\sum_{k>0}\epsilon_k(g)\left(\hat\gamma^\dagger_k(t)\hat\gamma_k(t)+\hat\gamma^\dagger_{-k}(t)\hat\gamma_{-k}(t)-1\right)$,
where $\hat\varphi_k=(\hat c_k,\hat c^\dagger_{-k})$ contains the creation and annihilation operators and the momentum is quantized as $k=\frac{\pi}{L},\,\frac{3\pi}{L},\,\dots,\,\pi-\frac{\pi}{L}$. The momentum space Hamiltonian associated with each mode is diagonalized by the Bogoliubov operators $\hat\gamma_k(t)$. The corresponding eigenenergies $\epsilon_k(g)$ exhibit avoided crossings in which pairs of quasiparticle excitations in modes $(-k,k)$ are generated with Landau-Zener (LZ) transition probabilities $p_k$, under a slow linear ramping of the form $g(t)=g(0)+t/\tau_Q$ \cite{Dziarmaga05,Cui2016}.
For processes terminating at $g(\tau_Q)=0$, the average number of pairs of defects (kinks) scales as $\kappa_1=\sum_{k>0}\langle\hat\gamma^\dagger_k(\tau_Q)\hat\gamma_k(\tau_Q)\rangle\sim\tau^{-1/2}_Q$, in agreement with the KZ mechanism for $z=\nu=1$ \cite{Damski05,Dziarmaga05,Zurek05,Dziarmaga2010}.
More generally, the associated number operator for defect pairs can be expressed in momentum space as $\hat N=\sum_{k>0}\gamma^\dagger_k(\epsilon_f)\gamma_k(\epsilon_f)$ for a given final quench depth, measured from the critical point, $\epsilon_f=g_f-g_c$. The average number of defect pairs  is $\kappa_1=\langle\hat N\rangle$. At fast quenches across $g_c$, this average scales linearly with the quench depth,  $\kappa_1 \sim \epsilon_f$, in agreement with the universal fast-quench scaling for $d=\nu=1$ \cite{Huabi2023FastQuench}. 

A characterization of nonequilibrium dynamics beyond KZ is provided by the full counting statistics of defects  \cite{Cincio07,delcampo18,Cui2020,Bando20,Subires22,King22,Gherardini24,King25,Kiss2025,visuri2025,Grabarits_FastCD_2025}. The probability of observing an eigenvalue $n$ of $\hat N$, $P(n)=\langle \delta (\hat N-n)\rangle$, can be characterized by cumulants, defined via the formal expansion $\log\langle e^{i\theta \hat N}\rangle =\sum_{q=1}^\infty\kappa_q(i\theta)^q/q!$. The distribution of the number of defect pairs $P(n)$ on which we focus is simply related to the distribution of the total number of defects \cite{Bando20}, further characterized in \cite{SM}. Cumulants are functions of $p_k$ and provided that $\sum_kp_k\gg \sum_kp_k^2$, they are proportional to the average, $\kappa_q\propto\kappa_1$. 
For quenches from the paramagnetic phase to the ferromagnetic phase, all $\kappa_q$ share the KZ scaling, that is, $\kappa_q\propto\tau^{-1/2}_Q$ \cite{delcampo18}.

\begin{figure}
    \includegraphics[width=.8\columnwidth]{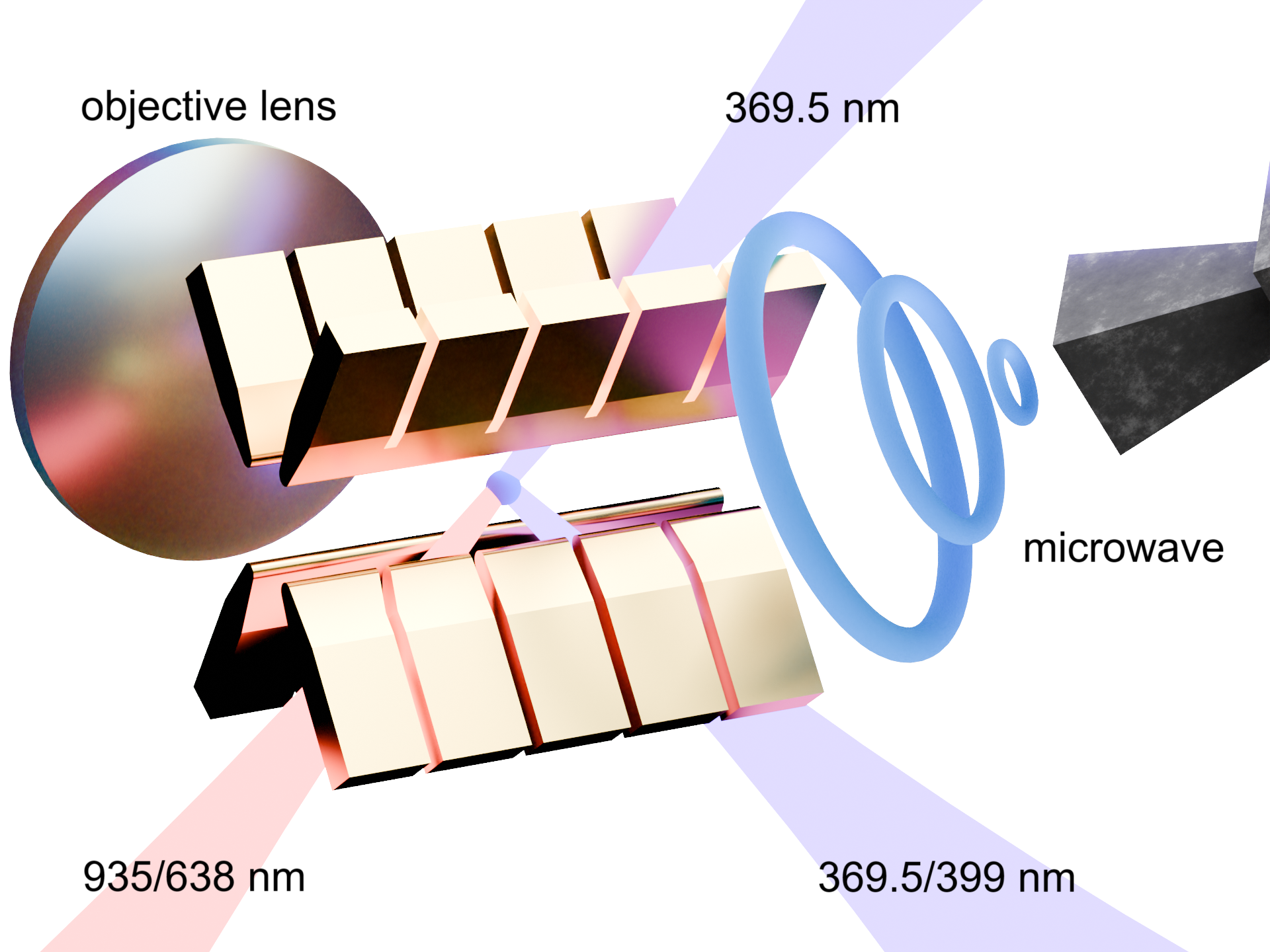}
    \caption{Scheme of the experimental setup for controlling a single $^{171}\mathrm{Yb}^+$ ion qubit. The ion is trapped at the center of a six-needle Paul trap (center), with surrounding control fields applied via laser beams and microwaves. Qubit initialization, manipulation, and readout are achieved using lasers at $369.5$ nm, $399$ nm, $638$ nm, and $935$ nm, focused by a high-NA (numerical aperture) objective lens (left). Microwave control is applied via near-resonant radiation around $12.442$ GHz (right), with ring-shaped antenna structures indicating the drive field. The labeled laser wavelengths correspond to Doppler cooling, repumping, as well as state preparation and readout transitions.}
    \label{fig:exp_setup}
\end{figure}

Measurement of defect statistics using real spins in quantum computers is practically infeasible due to the complexity of the spin-operator representation for any $g_f\neq0$. 
In contrast, trapped-ion simulators \cite{Cui2016,Cui2020} provide a natural platform for directly accessing the exact defect number, enabling the first experimental investigation of fast-quench defect statistics. 
We explore quantum quenches of varying depth that begin near criticality and characterize the associated full counting statistics of defects. 
To this end, each independent TLS is modeled by an ion trap qubit implemented with a $^{171}$Yb$^+$ ion confined in a Paul trap consisting of six needles placed on two perpendicular planes, as shown in Fig.~\ref{fig:exp_setup}. The hyperfine clock transition in the ground state $S_{1/2}$ manifold is chosen to realize the qubit, with energy levels denoted by $\lvert 0\rangle=\lvert F=0,\,m_F=0\rangle$ and $\lvert1\rangle\equiv\lvert F=1,\,m_F=0\rangle$.

\begin{figure}
    \includegraphics[width=.95\columnwidth]{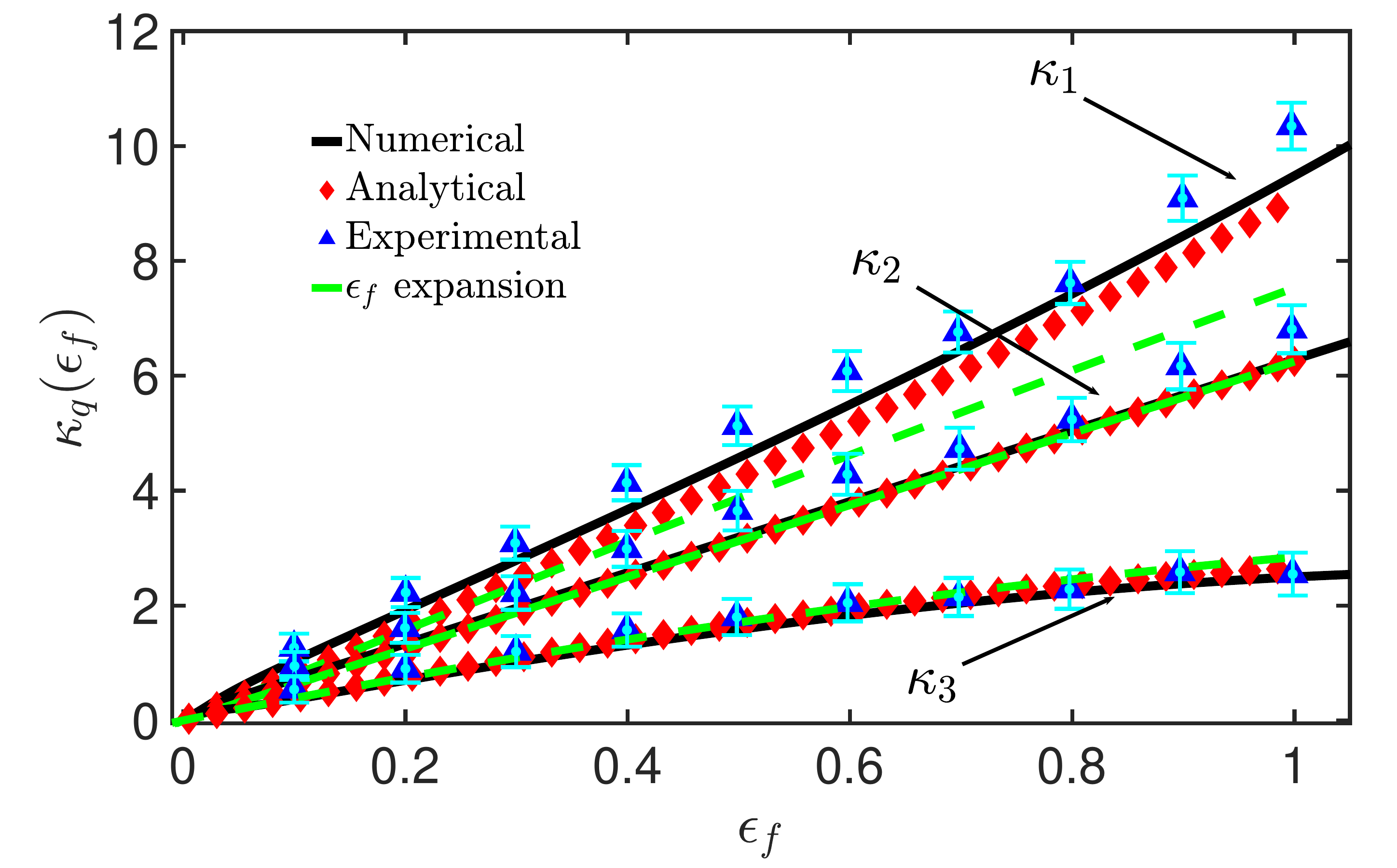}
    \caption{First three defect pair cumulants as a function of the quench depth measured from the $\epsilon_f = 0$ to $\epsilon_f = 1$ approached from near the critical point $g_i = -1.01$. Experimentally, we use the average of the data in the fast quench region to fit the results of the numerical simulations. Numerical findings (black dots) perfectly match the analytical approximations (red circles and green dashed line) and are also in agreement with the experimental results (blue triangles) for the variance and the third cumulant. The average is only recovered in the close neighborhood of the critical point.}
    \label{fig: cumulants_experimental}    
\end{figure}

\emph{Universal defect scaling.} 
In the experimental realization of the fast-quench dynamics, the initial transverse field is set close to the critical point, $g_i = -1.01$, choosing  $\tau_Q = 0.07$ 
to ensure that the time evolution lies deep within the fast-quench regime. The quench depth is varied in the range $g_f \in [-1, 0]$ for a system size $N = 100$. We parameterize the final quench depth by the distance from the critical point $\epsilon_f=g_f+1$, with
$\epsilon_f \in [0, 1]$. As shown in Fig.~\ref{fig: cumulants_experimental}, the average and variance of $P(n)$ obtained experimentally are proportional to each other and grow linearly with high precision. In contrast, the third cumulant exhibits a slower increase, consistent with stretched power-law behavior.

 Next, we carry out an exact calculation of the first three cumulants of $P(n)$ for an initial coupling $g_i = -1$, and also shed light on their analytical properties by expansion in terms of the final quench depth $g_f$ up to the second order.
The fast quench limit admits a precise analytical understanding via the overlap between the initial ground state $\ket{\mathrm{GS}_k(g_i = -1)}$ and the final excited state $\ket{\mathrm{ES}_k(g_f)}$ in a given $k$-th momentum mode, $p_k=\lvert\langle\mathrm GS_k(g_i)\vert \mathrm{ES}_k(g_f)\rangle\rvert^2$ \ref{app: diag}. First, the low-order $\epsilon_f$ expansion of the average is tested. Although it precisely matches the numerical results in the neighborhood of the critical point, slight deviations arise from it. In contrast, the second cumulant is exactly linear in $\epsilon_f$ for arbitrary quench depths,
\begin{eqnarray}
    \kappa_1\approx&&\frac{L}{4\pi}\left[\epsilon_f+\frac{\pi-4}{8}\epsilon^2_f-\frac{3\pi-10}{24}\epsilon^3_f-\frac{15\pi-48}{128}\epsilon^4_f\right],\nonumber\\
    &&+O(\epsilon^5_f)\\
    \kappa_2=&&\frac{L}{16}\epsilon_f.
\end{eqnarray}
The comparison of the exact analytical, numerical, and experimental results is shown in Fig.~\ref{fig: cumulants_experimental}. Although the variance perfectly follows the linear slope, the average shows small deviations away from the critical point.
To illustrate the validity regime for the fast-quench scaling of the experimental results, the numerical cumulants were obtained for the same number of modes $N=100$ and initial transverse field $g_i=-1.01$, using a shorter quench time $\tau_Q = 0.01$, allowing a direct comparison with the exact sudden limit with $\tau_Q = 0$. We further show how a second-order expansion $g_f=0$ with $g_i=-1.01$ provides slightly better results for the average over a broader range in \cite{SM}.

Similarly to the average, the third cumulant admits a rather involved closed expression, as shown in \cite{SM}. Yet, at variance with the average, the second-order term in its $\epsilon_f$ expansion is no longer negligible. This leads to a noticeable suppression in the skewness correction,
\begin{eqnarray}
    \kappa_3=&&\frac{L}{32\pi}\left[\epsilon_f-\frac{\pi-2}{128}\epsilon^2_f-\frac{3\pi-10}{192}\epsilon^3_f\right]+O(\epsilon^4_f)
\end{eqnarray}
where the coefficient of the quadratic term exceeds that of the linear one. This behavior, along with the agreement between the analytical, numerical, and experimental results, is also illustrated in Fig.~\ref{fig: cumulants_experimental}. For better distinguishability, the plot shows the cumulants of the number of pairs of defects. Additional details of the exact second-order expansion of the excitation probabilities at $g_i=-1.01$ and the corresponding cumulants as a function of $g_f$ around $g_f=0$ are shown in \cite{SM}. 

\begin{figure}[t]
    \includegraphics[width=.9\columnwidth]{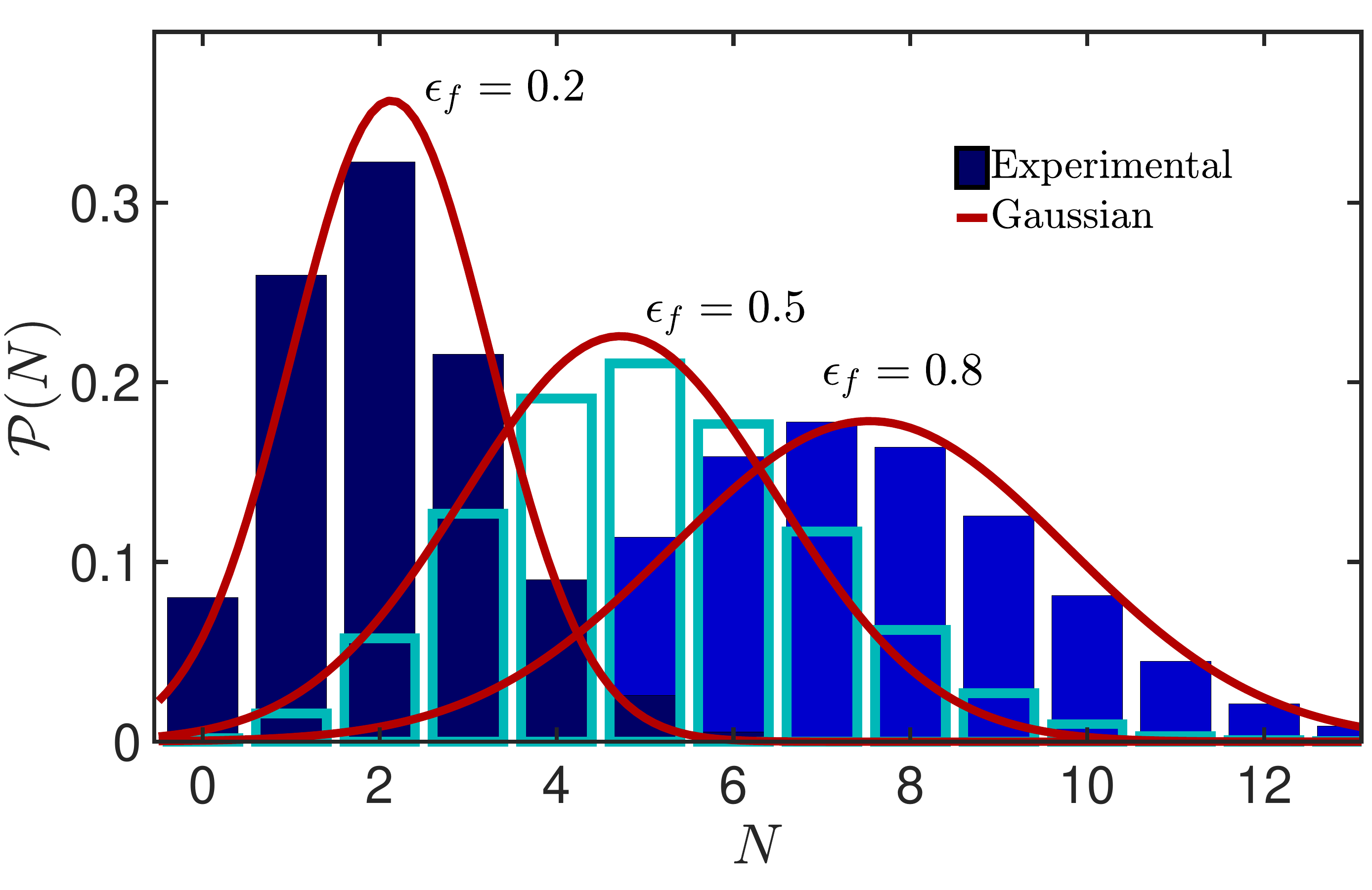}     
    \caption{Histograms of the defect pair number in sudden quenches stopping at different depths $\epsilon_f$. With the initial value of the experimental results, $g_i=-1.01$, precise matching is displayed with the Gaussian limiting distribution.}
    \label{fig:KinkStat}
\end{figure}

As a significant consequence, the defect pair number distribution remains consistently sub-Poissonian~\cite{Grabarits_Arbitrary2025,Grabarits_FastCD_2025,visuri2025}, both in terms of variance and skewness, $\kappa_2 / \kappa_1 < 1$ and $\kappa_3 / \kappa_1 < 1$. In contrast to conventional critical dynamics \cite{delcampo18,Cui2020,Bando20,King22}, the cumulants of the defect number become super-Poissonian, $2\kappa_2>\kappa_1,\,4\kappa_3>\kappa_1$, as detailed in \cite{SM}. We validate the analytical approximations by analyzing the statistics of defect pairs. In the thermodynamic limit, the transition probabilities $p_k$ become continuous functions of $k$, leading to a limiting distribution that is Gaussian. This distribution is fully characterized by the first two cumulants, $\kappa_1$ and $\kappa_2$, while the third cumulant, $\kappa_3$, quantifies the rate at which skewness corrections vanish. As a result, the limiting distributions exhibit behavior analogous to the KZ scaling regime with a paramagnetic initial state: both the peak position and the width scale proportionally. As shown in Fig.~\ref{fig:KinkStat}, the histograms of $P(n)$ experimentally obtained closely follow Gaussian distributions, with the average and variance determined from the experimental data.

\begin{figure}
    \includegraphics[width=.95\columnwidth]{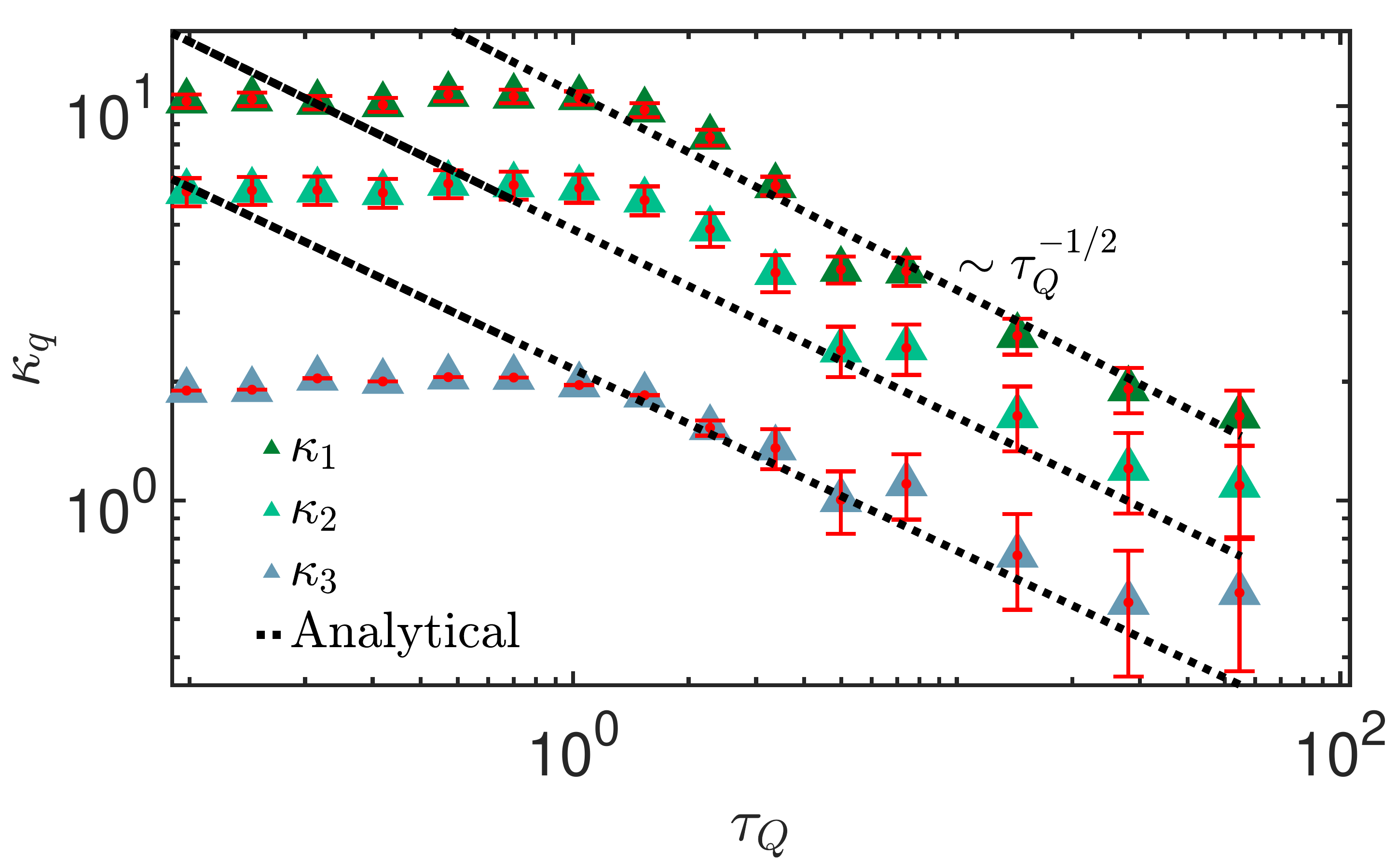}
    \caption{Kibble-Zurek power-law scaling for the first three defect pair cumulants and their breakdown at fast quenches, validated by the analytical approximations. Up to small deviations, precise matching is displayed for the overall power-law behavior ($N=100$ and $g_i=-1.01$ for both the experimental and numerical results).}
    \label{fig:Cumulants_slow_experimental}
\end{figure}

    \emph{Slow-driving universality.} 
    The trapped-ion simulator also enables experimental access to the scaling of the first three cumulants in the slow-driving limit, $\tau_Q\gtrsim1$. These quantum simulations correspond to the same initial transverse field near the critical point ($g_i = -1.01$), while keeping the final quench depth fixed, $g_f=0$, and varying $\tau_Q$ from the onset of the fast-quench limit to near-adiabatic dynamics. As shown in Fig.~\ref{fig:Cumulants_slow_experimental}, the cumulants exhibit a characteristic crossover from a fast-quench plateau to a power-law decay, $\sim \tau_Q^{-1/2}$, in agreement with KZ scaling behavior for larger initial $g_i$.

    To complement these experimental results, we analyze the dynamics using approximate analytical and numerical approaches. Although the exact time-evolved wavefunction for each mode can be expressed in terms of parabolic cylinder functions~\cite{Grabarits_Arbitrary2025}, determining their correct combination and accessing their limiting forms in the slow-driving regime remains analytically intractable. However, since the time-dependent Schr\"odinger equation in the leading order depends on the combination of $k^2 \tau_Q$, the transition probabilities are expected to inherit this scaling behavior. This is further refined through minimal numerical input, which reveals a characteristic exponential suppression, $p_k \approx e^{-\pi k\sqrt{3\tau_Q/2}}$. The numerical validation of this ansatz is presented in Appendix~\ref{app: p_k_app}.

As in the KZ scaling scenario with a large initial $g_i$, the dominant contribution arises from the low-momentum sector, which directly leads to the same scaling $\tau_Q^{-1/2}$ for all cumulants, albeit with different coefficients. In sharp contrast to the general sub-Poissonian behavior for large initial $g_i$, both in the KZ scaling regime and for fast quenches with varying $g_f$, the cumulant ratios increase. In particular, within the exponential approximation, 
 one finds the approximate cumulant scaling $\kappa_1\approx L(6\pi^4\tau_Q)^{-1/2},\,\kappa_2\approx L(6\pi^4\tau_Q)^{-1/2}/2,\,\kappa_3\approx L(6\,\pi^4\tau_Q)^{-1/2}/3$ implying that both the variance and the skewness of the defect pair number remain sub-Poissonian, $\kappa_2< \kappa_1$, $\kappa_3 < \kappa_1$.
\begin{figure}[t]
    \includegraphics[width=.9\columnwidth]{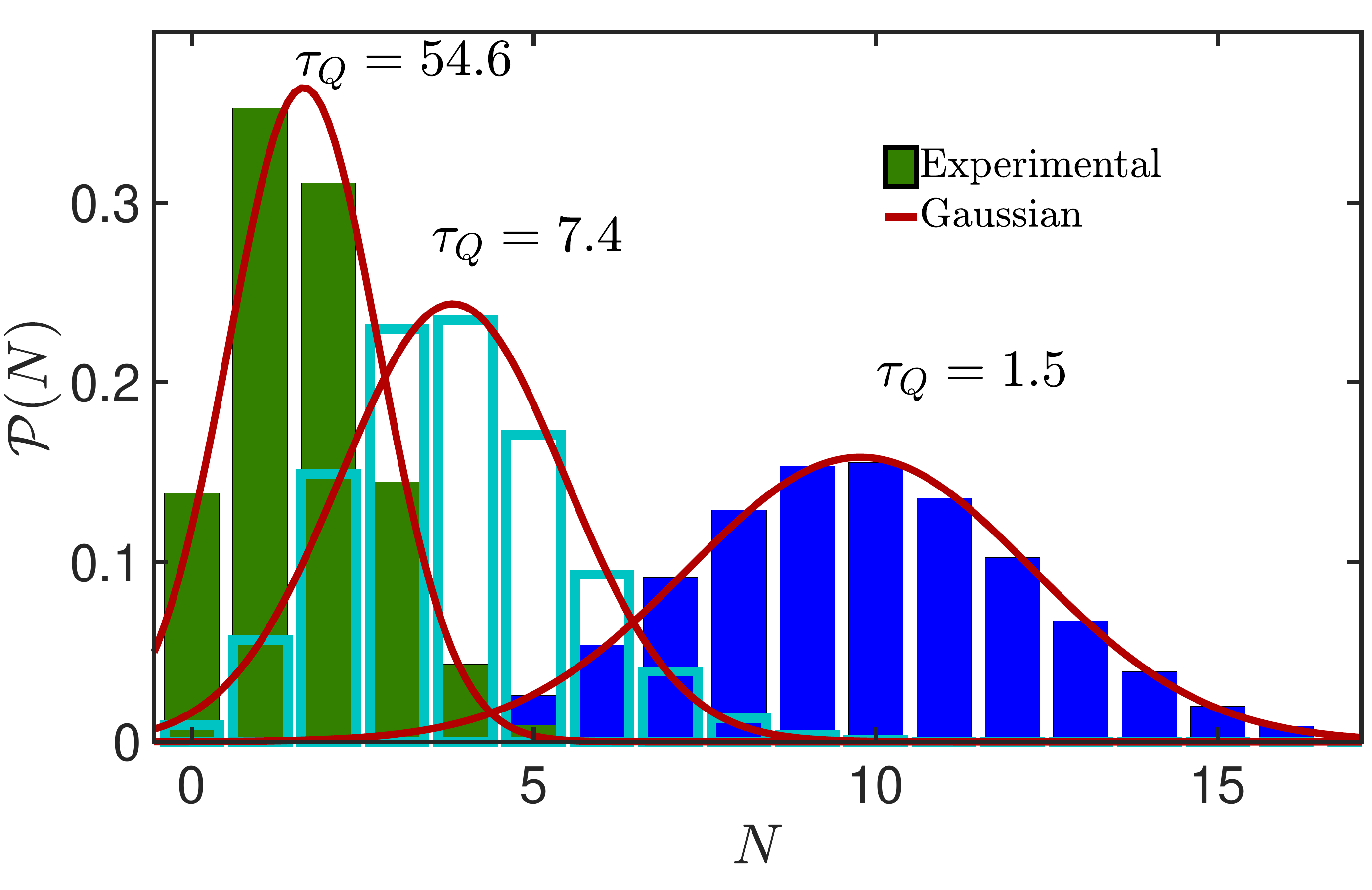}
    \caption{Defect pair number statistics for slow driving. For smaller driving times, slight deviations are observed, while good matching is found inside the power-law scaling regime with the limiting Gaussian distribution. ($\tau_Q=1.5,\,7.4,\,54.6$, $N=100$ and $g_i=-1.01$). }
    \label{fig:KinkStat_slow}    
\end{figure}
These analytical approximations, up to a small constant prefactor,  closely match the experimental data, as illustrated in Fig.~\ref{fig:Cumulants_slow_experimental}.
Remarkably, the defect number cumulants display super-Poissonian behavior in the slow driving regime as well, as detailed in \cite{SM}. In particular,
beyond the leading-order exponential approximations, small deviations break this equivalence, leading to an overall super-Poissonian character for the variance, $\kappa_1 < 2\kappa_2$, while the inequality $4\kappa_3>\kappa_1$ is readily captured by the exponential approximation. These subtle corrections are most clearly revealed in the cumulant ratios, as detailed in \cite{SM}.

Finally, the theoretical predictions and experimental results for the defect statistics are presented in Fig.~\ref{fig:KinkStat_slow}. Within the power-law scaling regime, we observe excellent agreement between numerical simulations and analytical predictions. At shorter driving times, around $\tau_Q \sim \mathcal{O}(1)$, a fast-quench plateau emerges, leading to minor deviations; however, the general Gaussian character of the distribution remains intact.

\emph{Conclusions.} 
We have reported the experimental verification of universal defect statistics in fast quantum quenches from a critical point, using a trapped-ion quantum simulator. Beyond the experimental observations, our study provides new theoretical insights into the dynamics of quantum phase transitions. Remarkably, Gaussian defect statistics and the linear scaling of the first two cumulants with quench depth remain robust even when the quench is initiated from the critical point, with the third cumulant exhibiting a slower growth due to quadratic suppression. In the slow-driving regime, universal  signatures beyond KZ scaling persist, despite the quenches starting near criticality.
These findings deepen our understanding of nonequilibrium quantum critical dynamics and establish quench-depth scaling as a stringent benchmark for quantum simulators and quantum computing platforms.

{\it Acknowledgements.} 
This project was supported by the Luxembourg National Research Fund (FNR Grant Nos. 17132054 and 17132O60), the National Key Research and Development Program of China (Grant No. 2024YFA1409403), the National Natural Science Foundation of China (Grant  No. 11275233, No. 11734015, and No. 12204455) and the Innovation Program for Quantum Science and Technology (Grant No. 2021ZD0301604 and No. 2021ZD0301200). 

\bibliography{references}

\begin{thebibliography}{44}
\bibitem [{\citenamefont {url will be inserted~by publisher}()}]{SM}%
  \BibitemOpen
  \bibfield  {author} {\bibinfo {author} {\bibnamefont {url will be inserted~by publisher}},\ }\href@noop {} {\ }\BibitemShut {NoStop}%
\end{thebibliography}

\newpage \appendix
\clearpage
\widetext

\section{TFQIM diagonalization}\label{app: diag}
With periodic boundary conditions, $\hat \sigma^z_1=\hat\sigma^z_{N+1}$, the TFQIM is translationally invariant and can be diagonalized using standard techniques. By performing a Jordan-Wigner transformation and a subsequent Fourier-mode decomposition, it can be written in the form $H=\sum_{k>0}H_k$ where the single-mode Hamiltonian $H_k=\hat\varphi^\dagger_k\left[(g-\cos k)\sigma^z+\sin k\sigma^x\right]\hat\varphi_k$ as defined in the main text has eigenvalues $\epsilon_k(g)=2\sqrt{1+g^2-2g\cos k}$.
The ground state at an initial arbitrary $g_i$ is given by
\begin{equation}
\ket{\mathrm{GS}_k(g_i=-1)} = \frac{(g_i-\cos k-\sqrt{1+g^2_i-2g_i\cos k} ,\sin k)^T}{\sqrt{\sin^2k+\left(g_i-\cos k-\sqrt{1+g^2_i-2g_i\cos k}\right)^2}}.
\end{equation}

The excited state in each mode for arbitrary $g_f$ reads
\begin{equation}
\ket{\mathrm{ES}_k(g_f)} = \frac{(g_f-\cos k+\sqrt{1+g^2_f-2g_f\cos k} ,\sin k)^T}{\sqrt{\sin^2k+\left(g_f-\cos k+\sqrt{1+g^2_f-2g_f\cos k}\right)^2}}.
\end{equation}
From this, the general expression for the excitation probabilities is given by
\begin{eqnarray}
    \label{eq:def_pk_fi}
    p^{fi}_k &=& \big |\braket{\mathrm{ES}_k(g_f)}{\mathrm{GS}_k(g_i)} \big|^2\\
    &=&  \frac{1}{\sin^2k+\left(g_f-\cos k+\sqrt{1+g^2_f-2g_f\cos k}\right)^2}\frac{1}{\sin^2k+\left(g_i-\cos k-\sqrt{1+g^2_i-2g_i\cos k}\right)^2}\nonumber\\
    &&\times\Big[\sin^4k+2\sin^2k\left(g_i-\cos k-\sqrt{1+g^2_i-2g_i\cos k}\right)\left(g_f-\cos k+\sqrt{1+g^2_f-2g_f\cos k}\right)\nonumber\\
    &&+\left(g_i-\cos k-\sqrt{1+g^2_i-2g_i\cos k}\right)^2\left(g_f-\cos k+\sqrt{1+g^2_f-2g_f\cos k}\right)^2\Big]\nonumber.
    \end{eqnarray}
    \section{Derivation of the fast quench cumulants}\label{app:cumulants}
    In this section, we derive the formulas for the exact fast-quench cumulants and their expansion around both $g_f=0$ and $\epsilon_f=0$.
    Starting from the critical point, $g_i=-1$, it can be shown that the ground state can be expressed as 
    \begin{eqnarray}
        \lvert\mathrm{GS}_k(g_i=-1)\rangle=
        \begin{pmatrix}
        -\cos(k/4)\\
        \sin(k/4)
        \end{pmatrix},
    \end{eqnarray}
    with eigenvalue of $-2\cos(k/2)$.
    and the exact excitation probabilities read 
\begin{eqnarray}
    \label{eq:pk_fi_zerothorder}
    p_k&=&\lvert\braket{\mathrm{ES}_k(g_f)}{\mathrm{GS}_k(g_i=-1)}\rvert^2\\
    &=&\frac{\sin^2 k\sin^2\frac{k}{4}+\cos^2\frac{k}{4}\left(g_f-\cos k+\sqrt{1+g^2_f-2g_f\cos k}\right)^2-\sin k\sin\frac{k}{2}\left(g_f-\cos k+\sqrt{1+g^2_f-2g_f\cos k}\right)}{\sin^2k+\left(g_f-\cos k+\sqrt{1+g^2_f-2g_f\cos k}\right)^2}.\nonumber
    \end{eqnarray}

We start with the average within the integral approximation $\kappa_1=2\sum_{k>0}\,p_k\approx(L/\pi)\int_0^\pi\mathrm dk\,p_k$. By straightforward computation, it can be seen that the excitation probabilities in Eq.~\eqref{eq:pk_fi_zerothorder} with $g_i=-1$ can be expressed in a convenient form as
\begin{eqnarray}
    p_k=\frac{\sin^2\frac{k}{4}}{2}\left(1-\frac{g_f-\cos k}{\sqrt{1+g^2_f-2g_f\cos k}}\right)+\frac{\cos^2\frac{k}{4}}{2}\left(1+\frac{g_f-\cos k}{\sqrt{1+g^2_f-2g_f\cos k}}\right)-\frac{1}{2}\frac{\sin k\sin\frac{k}{2}}{\sqrt{1+g^2_f-2g_f\cos k}}
\end{eqnarray}
As a result, the average can be expressed as
    \begin{eqnarray}
        \kappa_1&&\approx\frac{L}{2\pi}\int_0^\pi\mathrm dk\,\frac{\sin^2\frac{k}{4}}{2}\left(1-\frac{g_f-\cos k}{\sqrt{1+g^2_f-2g_f\cos k}}\right)+\frac{\cos^2\frac{k}{4}}{2}\left(1+\frac{g_f-\cos k}{\sqrt{1+g^2_f-2g_f\cos k}}\right)-\frac{1}{2}\frac{\sin k\sin\frac{k}{2}}{\sqrt{1+g^2_f-2g_f\cos k}}\nonumber\\
        &&=\frac{L}{4\pi}\int_0^\pi\mathrm dk\,\left[1+\frac{\cos\frac{k}{2}(g_f-\cos k)-\sin k\sin\frac{k}{2}}{\sqrt{1+g^2_f-2g_f\cos k}}\right]=\frac{L}{2}+\frac{L}{\sqrt{8\pi^2}}\int_{-1}^1\mathrm dt\,\frac{\sqrt{1+t}\,(g_f-t)-\sqrt{1-t}\sqrt{1-t^2}}{\sqrt{1-t^2}\sqrt{1+g^2_f-2g_ft}}\nonumber\\
        &&=\frac{L}{4}-\frac{L}{4\pi}(1-g_f)\frac{\arcsin\left(\frac{2\sqrt{-g_f}}{1-g_f}\right)}{\sqrt{-g_f}},\nonumber
    \end{eqnarray}
    valid for $g_f<0$. Expanding up to the fifth order, it becomes apparent that for $g_f\in[-1,0]$ the average is dominated by the linear order
    \begin{eqnarray}
        \kappa_1=\frac{L}{2\pi}\left(\frac{\pi-2}{2}+\frac{2}{3}g_f+\frac{2}{15}g^2_f+\frac{2}{35}g^3_f+\frac{2}{64}g^4_f+\frac{2}{99}g^5_f+\dots\right).
    \end{eqnarray}
    This expression is further expressed in terms of the distance from the critical point, $g_c=-1$, $\epsilon_f=1+g_f$, leading to
    \begin{eqnarray}
        \kappa_1\approx\frac{L}{4}-\frac{L}{4\pi}(2-\epsilon_f)\frac{\arcsin\left(\frac{2\sqrt{1-\epsilon_f}}{2-\epsilon_f}\right)}{\sqrt{1-\epsilon_f}}.
    \end{eqnarray}
    A more compact expansion scheme follows with $\epsilon_f$, yielding
    \begin{eqnarray}
        \kappa_1\approx \frac{L}{4\pi}\left[\epsilon_f+\frac{\pi-4}{8}\epsilon^2_f-\frac{3\pi-10}{24}\epsilon^3_f-\frac{3(5\pi-16)}{128}\epsilon^4_f-\frac{105\pi-332}{960}\epsilon^5_f+\dots\right].
    \end{eqnarray}

        For the variance, the individual Bernoulli variances can be written as
    \begin{eqnarray}
        p_k(1-p_k)=\frac{1}{4}\left[1-\frac{\left(\cos\frac{k}{2}(g_f-\cos k)-\sin k\sin\frac{k}{2}\right)^2}{1+g^2_f-2g_f\cos k}\right].
    \end{eqnarray}
    Thus, summing up the individual variances results in
    \begin{eqnarray}
        \kappa_2&&\approx\frac{L}{2\pi}\int_0^\pi\mathrm dk\,\frac{1}{4}\left[1-\frac{\left(\cos\frac{k}{2}(g_f-\cos k)-\sin k\sin\frac{k}{2}\right)^2}{1+g^2_f-2g_f\cos k}\right]=\frac{L}{2}-\frac{L}{4\pi}\int_{-1}^1\mathrm dt\,\frac{\left[\sqrt{1+t}(g_f-t)-\sqrt{1-t}\sqrt{1-t^2}\right]^2}{\sqrt{1-t^2}(1+t^2-2g_ft)}\\
        &&=\frac{L}{8}-\frac{L}{16\pi}(g_f-1)^2\int_{-1}^1\mathrm dt\,\frac{\sqrt{1+t}}{\sqrt{1-t}(1+g^2_f-2g_ft)}=\frac{L}{8}-\frac{L}{16}(g_f-1)^2\frac{1}{(1-g_f)}=\frac{L}{8}-\frac{L}{16}(1-g_f)=\frac{L}{16}+\frac{L}{16}g_f,\nonumber
    \end{eqnarray}
    which, in terms of the distance from the critical point, becomes
    \begin{eqnarray}
        \kappa_2\approx\frac{L}{16}\epsilon_f.
    \end{eqnarray}
    Finally, the third cumulant can also be obtained exactly, but in a rather involved form, similar to the average,
    \begin{eqnarray}
        p_k(1-p_k)(1-2p_k)=\frac{1}{4}\left[1-\frac{\left(\cos\frac{k}{2}(g_f-\cos k)-\sin k\sin\frac{k}{2}\right)^2}{1+g^2_f-2g_f\cos k}\right]\frac{\sin k\sin\frac{k}{2}-\cos\frac{k}{2}(g_f-\cos k)}{\sqrt{1+g^2_f-2g_f\cos k}},
    \end{eqnarray}
    which leads to the integral of
    \begin{eqnarray}
        \kappa_3&&\approx\frac{L}{2\pi}\int_0^\pi\mathrm dk\,\frac{1}{4}\left[\frac{\sin k\sin\frac{k}{2}-\cos\frac{k}{2}(g_f-\cos k)}{\sqrt{1+g^2_f-2g_f\cos k}}+\frac{\left(\cos\frac{k}{2}(g_f-\cos k)-\sin k\sin\frac{k}{2}\right)^3}{\left(1+g^2_f-2g_f\cos k\right)^{3/2}}\right]\\
        &&=\frac{L}{8\pi}\int_{-1}^1\mathrm dt\,\frac{g_f-1}{\sqrt2\sqrt{1-t}\sqrt{1+g^2_f-2g_ft}}+\frac{\left(\sqrt{1+t}(g_f-t)-\sqrt{1-t}\sqrt{1-t^2}\right)^3}{\sqrt{8}\sqrt{1-t^2}\left(1+g^2_f-2g_ft\right)^{3/2}}\nonumber\\
        &&=\frac{L}{8\pi}(1-g_f)\frac{\arcsin\left(\frac{2\sqrt{-g_f}}{1-g_f}\right)}{\sqrt{-g_f}}+\frac{L}{\sqrt{256\pi^2}}(g_f-1)^3\int_{-1}^1\mathrm dt\,\frac{1+t}{\sqrt{1-t}\left(1+g^2_f-2g_ft\right)^{3/2}}\nonumber\\
        &&=\frac{L}{8\pi}(1-g_f)\frac{\arcsin\left(\frac{2\sqrt{-g_f}}{1-g_f}\right)}{\sqrt{-g_f}}+\frac{L}{64\pi}\left[\frac{\arccos\left(\frac{g^2_f+6g_f+1}{(1-g_f)^2}\right)}{\sqrt{-g^3_f}}(g_f-1)^3+4\frac{1-g^2_f}{g_f}\right].\\
    \end{eqnarray}
    The leading order expansion in terms of $g_f$ reads     
    \begin{eqnarray}
        \kappa_3\approx \frac{L}{8\pi}\left(\frac{2}{3}-\frac{4}{15}\,g_f-\frac{4}{21}\,g^2_f-\frac{4}{63}g^3_f-\frac{116}{3465}g^4_f-\frac{188}{9009}g^5_f+\dots\right),\nonumber\\
    \end{eqnarray}
    In terms of the distance for the critical point, the leading-order behavior takes the form of
    \begin{eqnarray}
        \kappa_3=\frac{L}{32\pi}\left[\epsilon_f-\frac{\pi-2}{128}\epsilon^2_f-\frac{3\pi-10}{192}\epsilon^3+\frac{48-15\pi}{1024\pi}\epsilon^4_f+\frac{332-105\pi}{7720}\epsilon^5_f+\dots\right].
    \end{eqnarray}
    
\section{Expansion at small initial transverse fields}\label{app: pfi_k_expansion}
\begin{figure*}
    \centering
    \begin{tabular}{c | c | c}
        \includegraphics[width=0.34\textwidth]{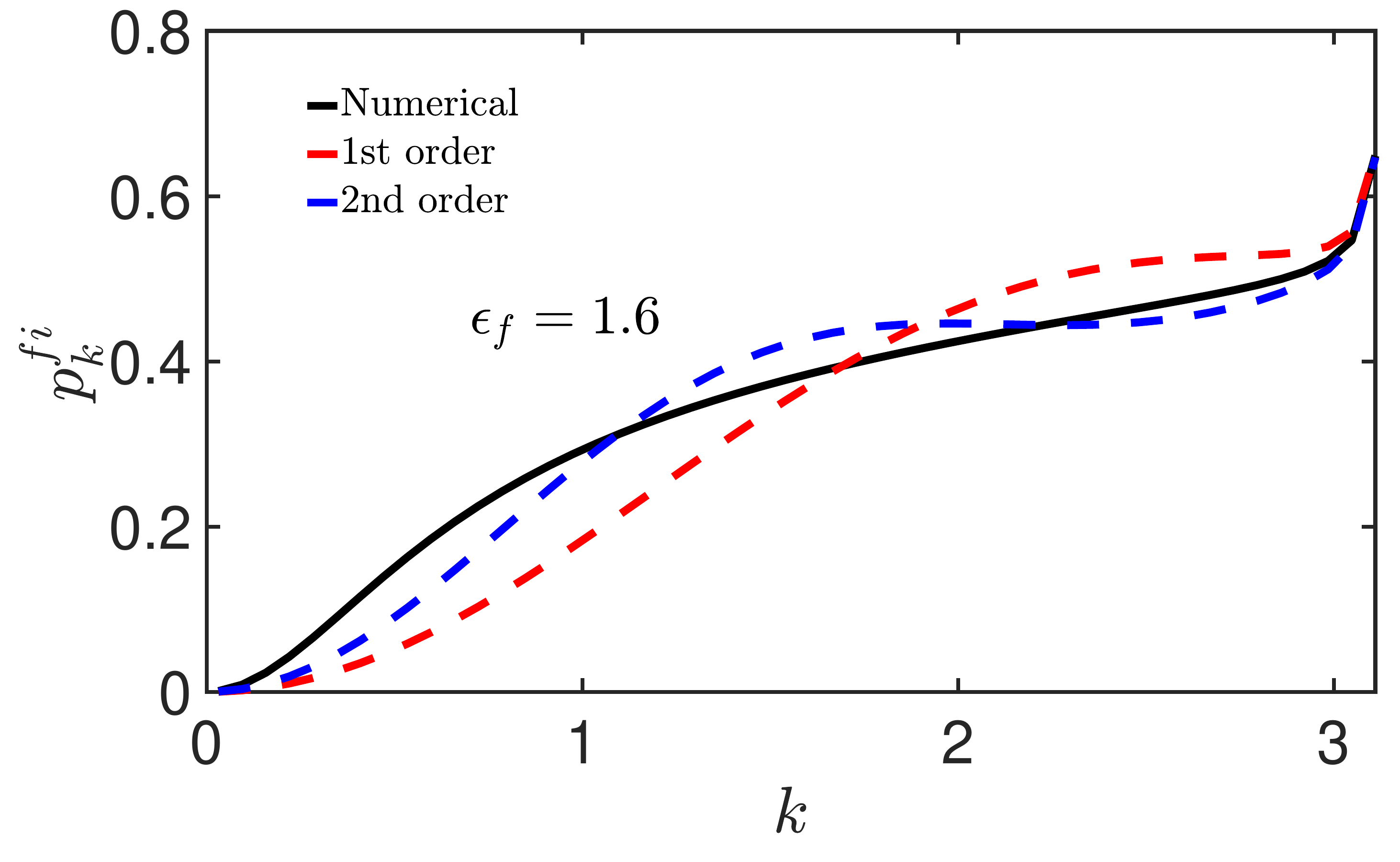} & 
        \includegraphics[width=0.34\textwidth]{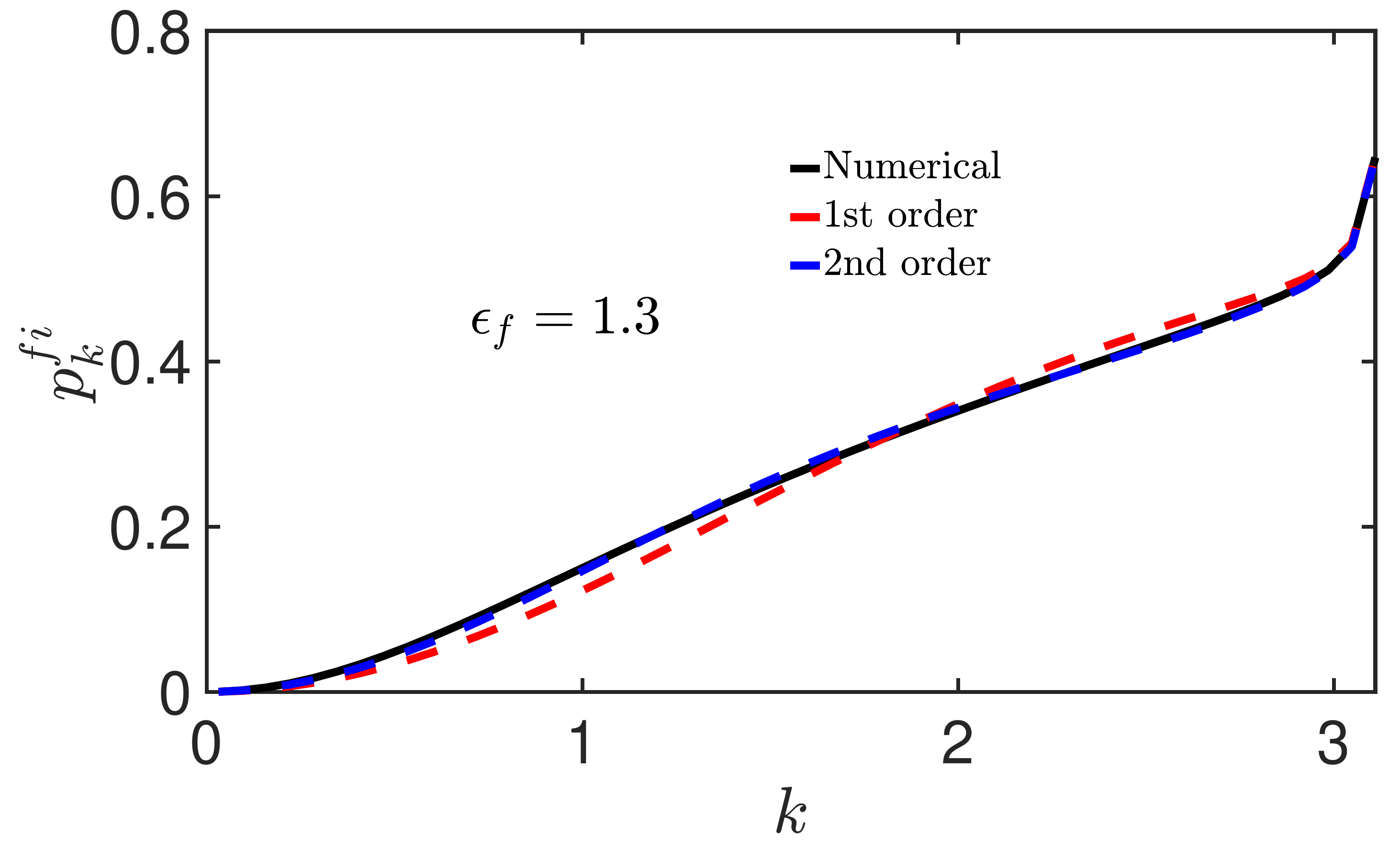} & 
        \includegraphics[width=0.34\textwidth]{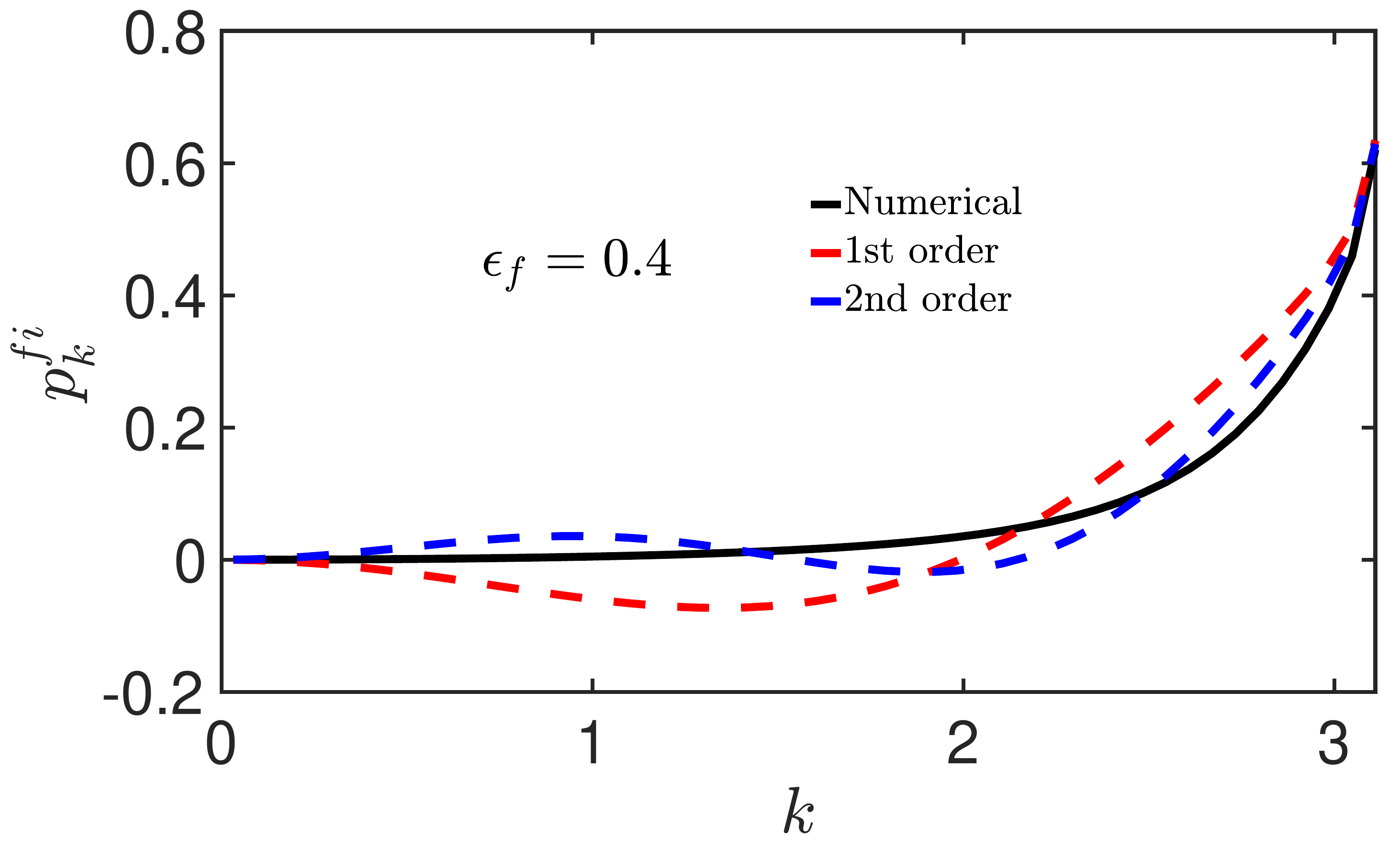} \\
    \end{tabular}
    \caption{Excitation probabilities up to second order in $g_f$ for the experimental value of the initial value of $g_i=-1.01$ as a function of $k$ for $\epsilon_f=1.6,\,1.3,\,0.4$. Approaching the critical points, the precision and accuracy deteriorate gradually.}
    \label{fig:pfi_k_approx}
\end{figure*}

\begin{figure}
\includegraphics[width=.5\textwidth]{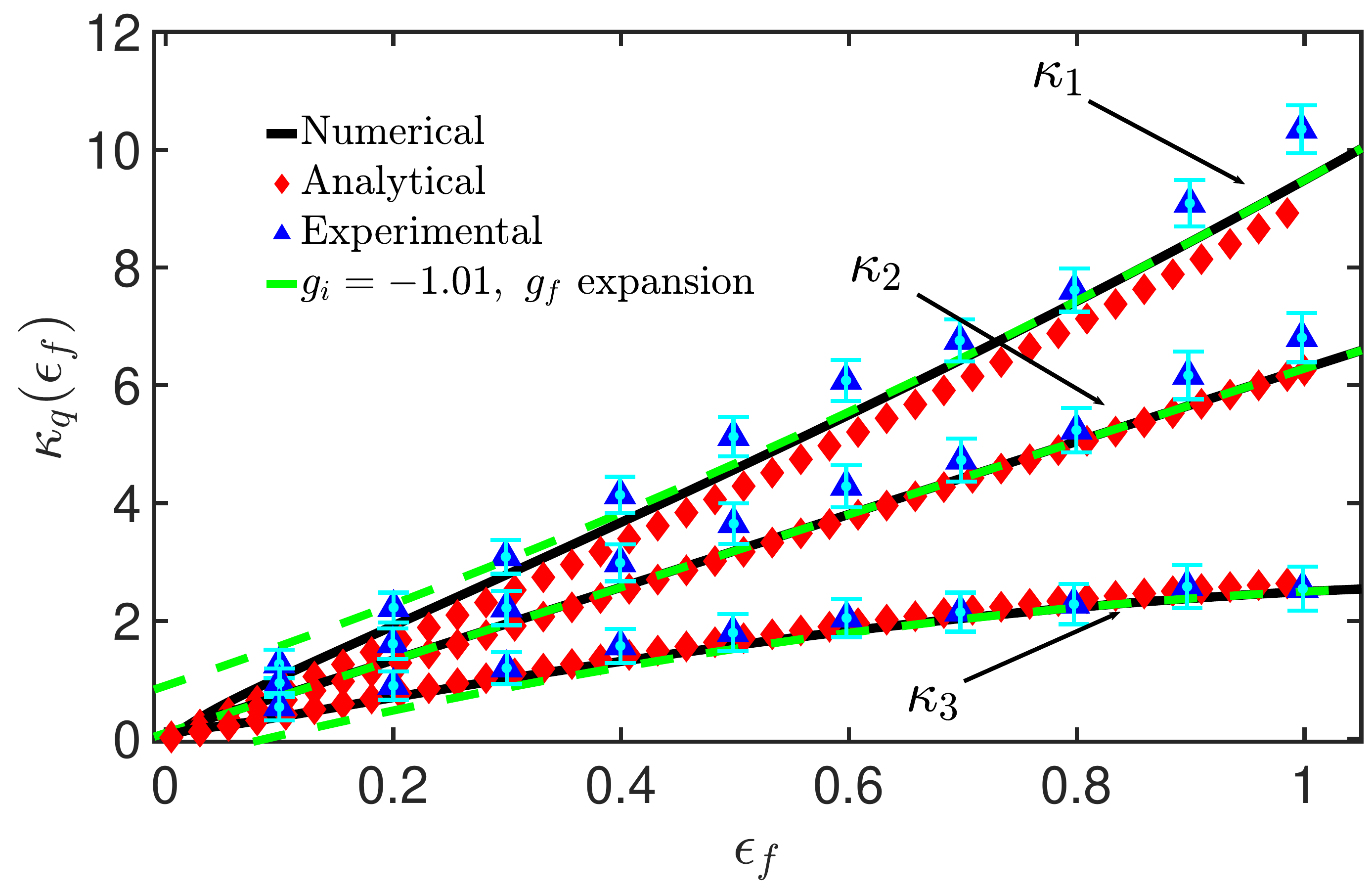}
\caption{First three cumulants as a function of the quench depth. A precise agreement is found for the second-order $g_f$ expansion with the experimental, numerical, and exact analytical results for $ g_i=-1$.}
\label{fig:g_f_expansion}
\end{figure}

In this appendix, we provide a second-order expansion of the excitation probabilities in Eq.~\eqref{eq:def_pk_fi}:

\begin{eqnarray}
    &&p^{fi}_k=\nonumber\\
    &&
\frac{4\sin^4\frac{k}{2} \left( -g_i + \cos k + \sqrt{1 + g^2_i - 2 g_i \cos k} \right)^2 + 
    4 \left( -g_i + \cos k + \sqrt{1 + g^2_i - 2 g_i \cos k} \right) \sin^2\frac{k}{2}\sin^2 k + 
    \sin^4 k
}{
    \left( (1 - \cos k)^2 + \sin k^2 \right) 
    \left( \left( g_i - \cos k - \sqrt{1 + g_i^2 - 2 g_i \cos k} \right)^2 + \sin^2 k \right)
}
\nonumber\\
&&+  g_f\Bigg( 
\frac{
    4\sin^4\frac{k}{2} \left( -g_i + \cos k + \sqrt{1 + g_i^2 - 2 g_i \cos k} \right)^2 + 
    4\sin^2\frac{k}{2} \left( -g_i + \cos k + \sqrt{1 + g_i^2 - 2 g_i \cos k} \right) \sin^2 k
}{
   \left( \left( g_i - \cos k - \sqrt{1 + g_i^2 - 2 g_i \cos k} \right)^2 + \sin k^2 \right)  \left( (1 - \cos k)^2 + \sin k^2 \right)
}
\nonumber\\
&&- 
\frac{
    32\sin^6\frac{k}{2} \left(\sin^2\frac{k}{2} \left( -g_i + \cos k + \sqrt{1 + g_i^2 - 2 g_i \cos k} \right)^2 + 
     \left( -g_i + \cos k + \sqrt{1 + g_i^2 - 2 g_i \cos k} \right) \sin k^2 + 
    \sin k^4 \right)
}{
   \left( \left( g_i - \cos k - \sqrt{1 + g_i^2 - 2 g_i \cos k} \right)^2 + \sin k^2 \right)  \left( (1 - \cos k)^2 + \sin k^2 \right)^2
}
\Bigg) 
\nonumber\\
&&
+ 
 g^2_f\Bigg( 
-\frac{
    32\sin^6\frac{k}{2} \left(2\sin^2\frac{k}{2} \left( -g_i + \cos k + \sqrt{1 + g_i^2 - 2 g_i \cos k} \right)^2 + 
    \left( -g_i + \cos k + \sqrt{1 + g_i^2 - 2 g_i \cos k} \right) \sin k^2\right)
}{
    \left( (1 - \cos k)^2 + \sin k^2 \right)^2\left(\left( g_i - \cos k - \sqrt{1 + g_i^2 - 2 g_i \cos k} \right)^2 + \sin k^2\right) 
} \nonumber\\
&&+ 
\frac{
    \left( 2 - 3 \cos k + \cos k^3 \right) \left( -g_i + \cos k + \sqrt{1 + g_i^2 - 2 g_i \cos k} \right)^2 + 
    (-1 + \cos k^2) \left( -g_i + \cos k + \sqrt{1 + g_i^2 - 2 g_i \cos k} \right) \sin k^2
}{
    \left( (1 - \cos k)^2 + \sin k^2 \right)\left(\left( g_i - \cos k - \sqrt{1 + g_i^2 - 2 g_i \cos k} \right)^2 + \sin k^2\right) 
}
\nonumber\\
&&+ 
\frac{
    \left( 4\sin^4\frac{k}{2} \left( -g_i + \cos k + \sqrt{1 + g_i^2 - 2 g_i \cos k} \right)^2 + 
    4\sin^2\frac{k}{2} \left( -g_i + \cos k + \sqrt{1 + g_i^2 - 2 g_i \cos k} \right) \sin k^2 + 
    \sin k^4 \right) 
}{
    \left(\left( g_i - \cos k - \sqrt{1 + g_i^2 - 2 g_i \cos k} \right)^2 + \sin k^2\right)\left((1 - \cos k)^2 + \sin k^2\right)
}\nonumber\\
&&\times\frac{\left( \frac{4 (-1 + \cos k)^4}{\left( (1 - \cos k)^2 + \sin k^2 \right)^2} - 
    \frac{2 - 3 \cos k + \cos k^3}{(1 - \cos k)^2 + \sin k^2} \right)}{\left(\left( g_i - \cos k - \sqrt{1 + g_i^2 - 2 g_i \cos k} \right)^2 + \sin k^2\right)\left((1 - \cos k)^2 + \sin k^2\right)}
\Bigg).
\end{eqnarray}
Despite the cumbersome expression, a precise matching is provided by the second-order expansion in terms of $g_f$ in the interval $g_f\in[-0.6,0.6]$, as displayed in Fig.~\ref{fig:pfi_k_approx}. The corresponding cumulants are also precisely reproduced up to $\epsilon_f\approx0.2$, as shown in Fig.~\ref{fig:g_f_expansion} with coefficients being close to the $g_i=-1$ expansion results. The numerical integrals of the $k$ dependent prefactors provide the prefactor for the cumulant expansions in the main text.

\pagebreak

\section{Demonstration of the approximate exponential excitation proabilities}\label{app: p_k_app}

\begin{figure}[h]
    \includegraphics[width=.5\textwidth]{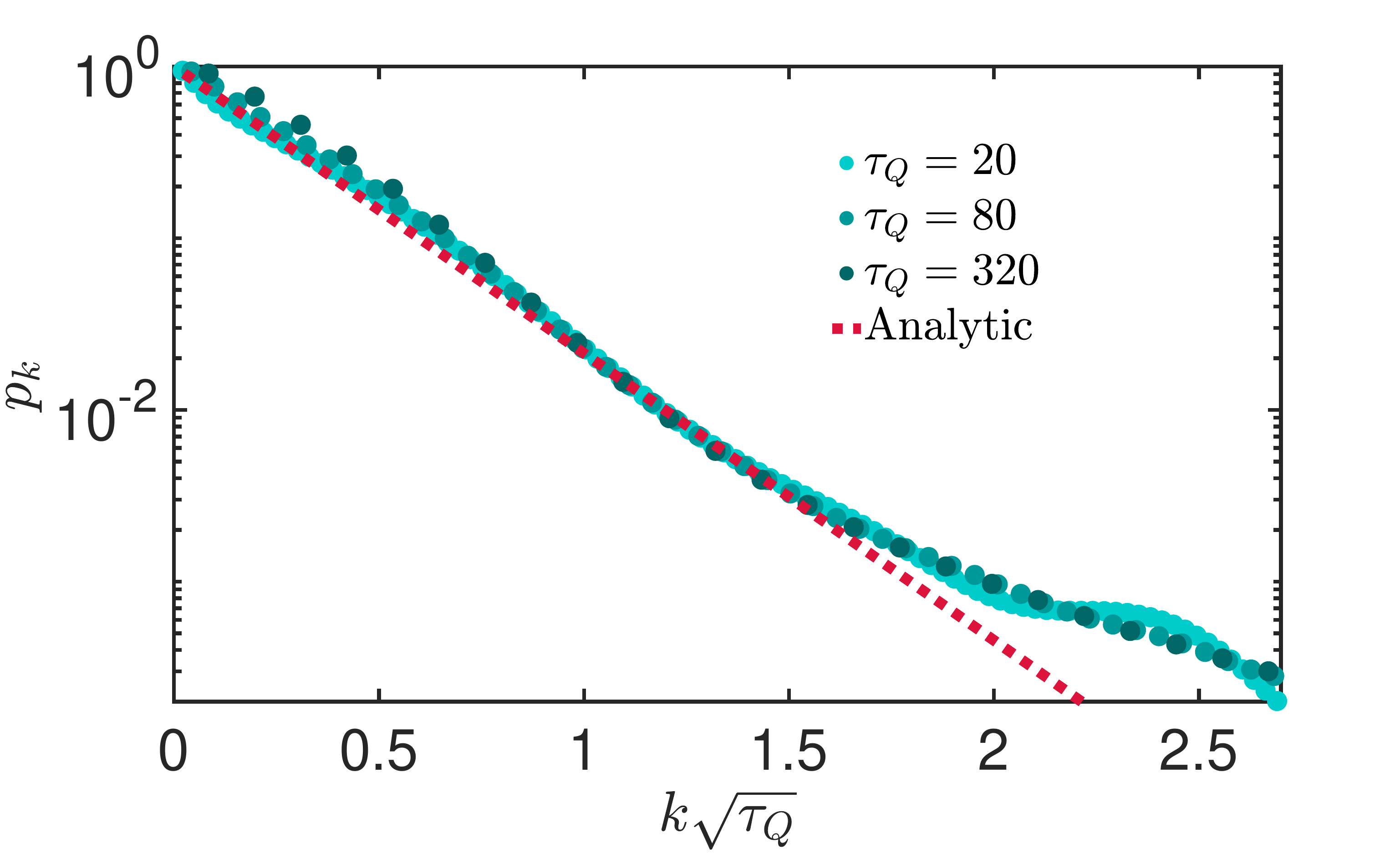}
    \caption{Excitation probabilities in the slow driving limit for different values of $\tau_Q$, exhibiting an approximate exponential decay, captured by the fitting in a semilog plot.}
    \label{fig:p_k_slow}    
\end{figure}

In this section, we show the numerical verification of the exponential ansatz for the excitation probabilities, $p_k\sim\,e^{-\pi\sqrt{3\tau_Q/2}\,k}$. In Fig.~\ref{fig:p_k_slow} we show that this ansatz provides a precise fitting for $p_k$, since for different values $\tau_Q$ an accurate scaling collapse is displayed as a function of $k\sqrt{\tau_Q}$ following an exponential decay for low momenta. Within its domain of validity, this approximation contains the largest contribution to defect production, $p_k\gtrsim5\times10^{-4}$.

             \begin{figure}
    \includegraphics[width=.5\columnwidth]{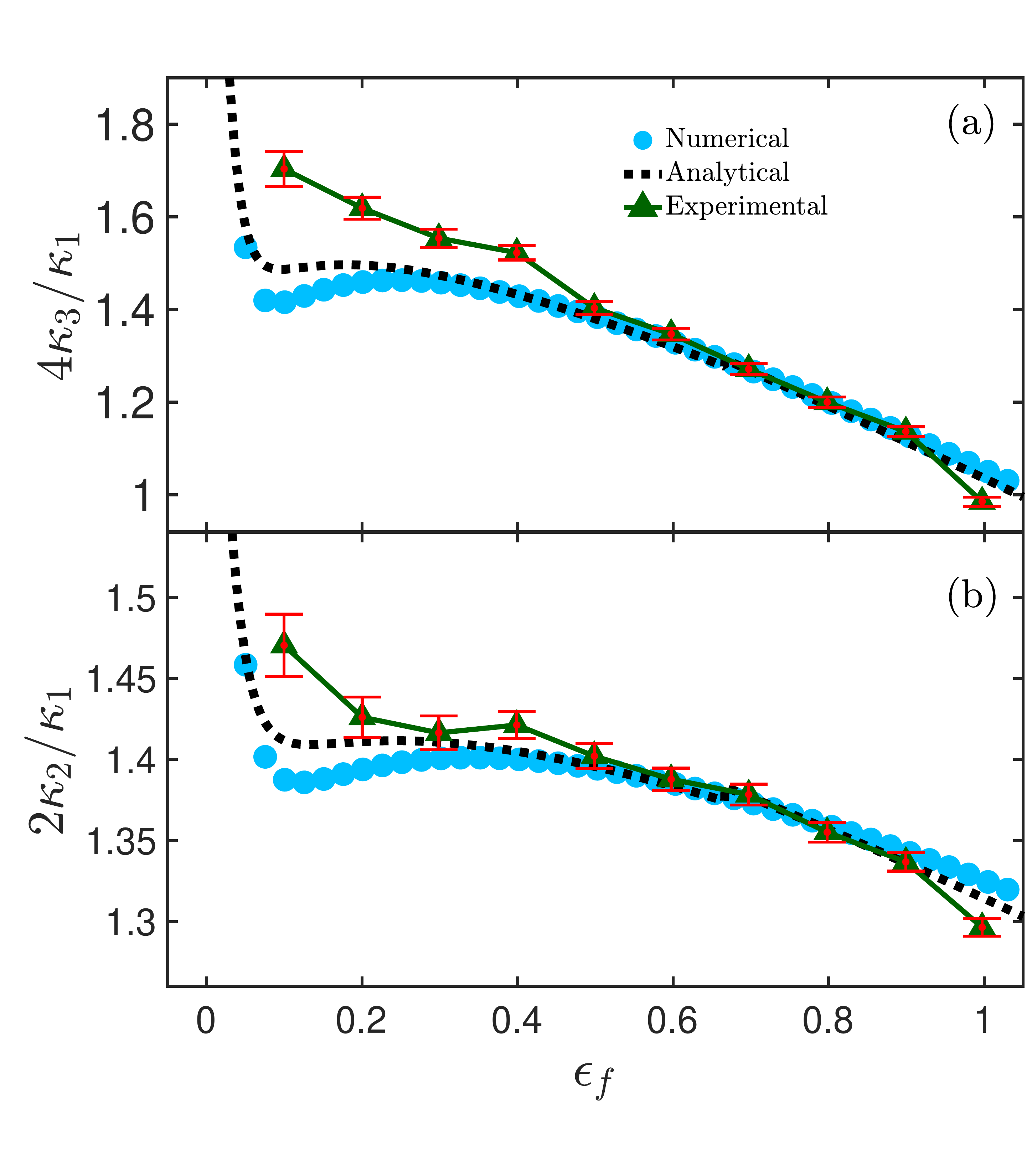}
    \caption{Cumulant ratios of defect numbers by experimental results, numerical simulations, and analytical formulas. In all three data sets, the defect number statistics is super-Poissonian for both the variance and the skewness. (a) Skewness ratio with super-Poissonian behavior converging to the Poissonian value at $\epsilon_f=1$. (b) Variance ratio, displaying a clear super-Poissonian behavior for the full range of $\epsilon_f$ ($\tau_Q=0.01$ and $N=100$). 
    }
    \label{fig:cumulant_ratios_experimental}    
\end{figure}

\section{Defect number cumulant ratios}\label{app: cumulants_ratios_slow}
In this section, we demonstrate that the cumulants of the defect numbers become super-Poissonian both in the fast and slow driving regimes. Using the exact results and leading-order expansions in the fast quench regime, it can be shown that the variance preserves its super-Poissonian behavior as  $2\kappa_1-4\kappa_2\approx \left(1/(2\pi)-1/4\right)\epsilon_f<0\Rightarrow 2\kappa_2/\kappa_1>1$, for the range of validity of the linear $\epsilon_f$ expansions. For larger values of $\epsilon_f$, the exact solution for $\kappa_1$ reveals the same super-Poissonian behavior. The skewness ratio follows straightforwardly from the analytical approximations, following a super-Poissonian curve for the dominant range of $\epsilon_f$. However, at the same time, it clearly converges to unity as $\epsilon_f=1$ is reached, implying a Poissonian character. The full dependence of these cumulant ratios on $\epsilon_f$ is presented in Fig.~\ref{fig:cumulant_ratios_experimental}.

\begin{figure}
    \includegraphics[width=.5\textwidth]{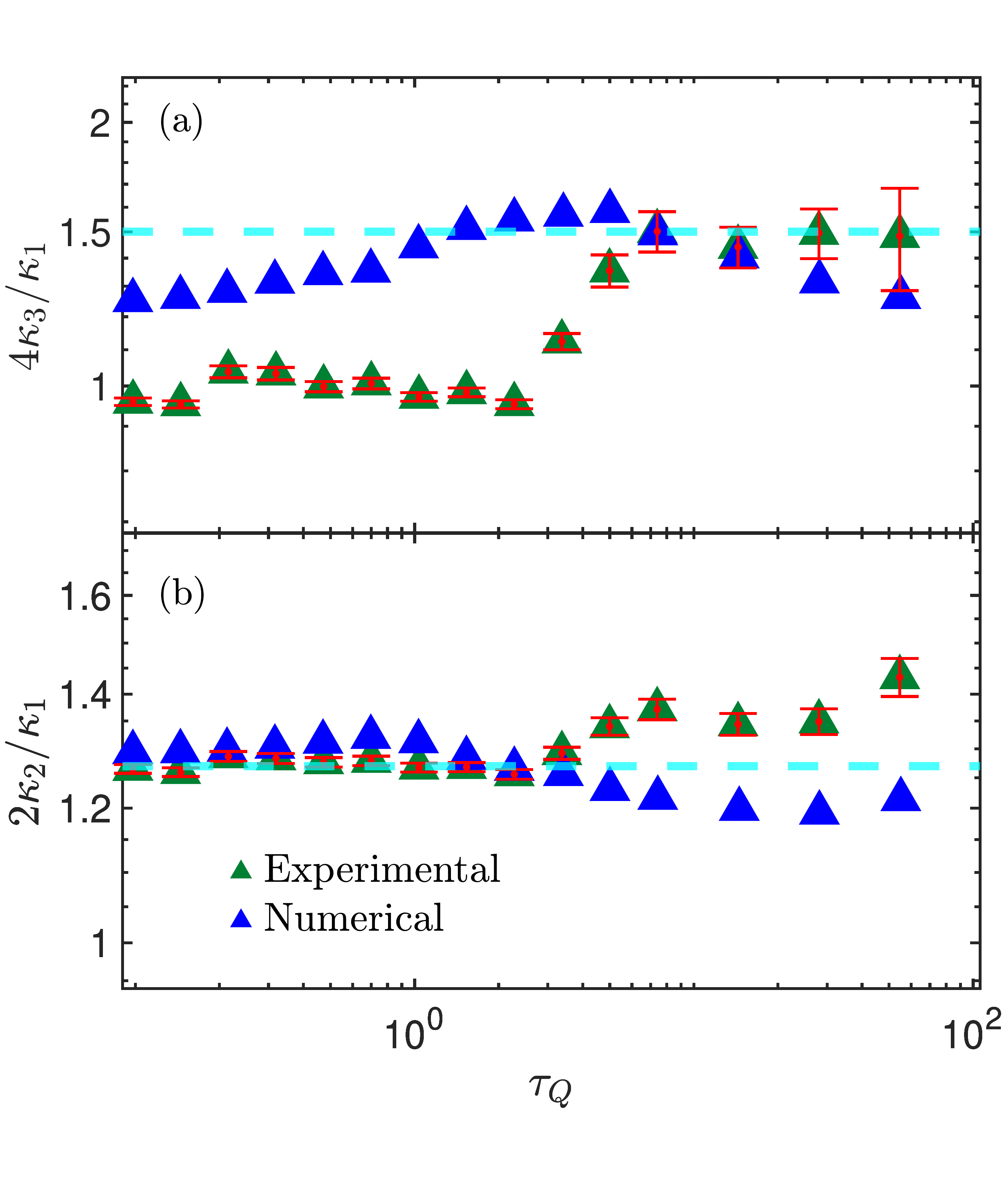}
    \caption{Defect number cumulant ratios with varying driving time and for $\epsilon_f=1$, exhibiting an approximately constant, super-Poissonian behavior. (a) The experimental results for the skewness ratio, besides small deviations from the numerical findings, closely match the predicted constant value, $\kappa_3/\kappa_1\approx1.5$ in the slow limit, while it converges to unity for fast quenches. (b) The variance ratio exhibits a slightly larger value than the one predicted by the exponential approximation, $2\kappa_2/\kappa_1\approx1.3$ ($\tau_Q=0.01$ and $N=100$).}
    \label{fig:cumulant_slow}    
\end{figure}

The $\tau_Q$ dependence of the cumulant ratios is shown in Fig.~\ref{fig:cumulant_slow}. Although the cumulants are well-captured by KZ-like power laws, constant-factor deviations become apparent in the cumulant ratios and eventually lead to an overall super-Poissonian behavior. As shown in Fig.~\ref{fig:cumulant_slow}(a), the ratio of skewness exhibits well the transition to super-Poissonian, closely matching the theoretical prediction for the exponential approximation, $4\kappa_3/\kappa_1\approx 1.5$. On the other hand, the variance ratio in panel (b) shows slight deviations from the predicted Poissonian behavior based on the exponential approximation, as upon increasing $\tau_Q$, it shifts to a slightly larger value, $2\kappa_2/\kappa_1\approx1.3$, tracing out a clear super-Poissonian behavior.

\end{document}